\documentclass[doublecol]{epl2}
\title{Selection of dynamical rules in spatial Prisoner's Dilemma games}
\shorttitle{Selection of dynamical rules in spatial Prisoner's Dilemma games}

\author{Gy\"orgy Szab\'o, Attila Szolnoki, and Jeromos Vukov}
\shortauthor{Gy\"orgy Szab\'o, Attila Szolnoki, and Jeromos Vukov}

\institute{Research Institute for Technical Physics and
Materials Science, P.O. Box 49, H-1525 Budapest, Hungary}


\abstract{We study co-evolutionary Prisoner's Dilemma games where each player can imitate both the strategy and imitation rule from a randomly chosen neighbor with a probability dependent on the payoff difference when the player's income is collected from games with the neighbors. The players, located on the sites of a two-dimensional lattice, follow unconditional cooperation or defection and use individual strategy adoption rule described by a parameter. If the system is started from a random initial state then the present co-evolutionary rule drives the system towards a state where only one evolutionary rule remains alive even in the coexistence of cooperative and defective behaviors. The final rule is related to the optimum providing the highest level of cooperation and affected by the topology of the connectivity structure.}

\begin{document}

\maketitle

The evolutionary game theory provides a general mathematical framework for the investigation of multi-agent systems used widely in biology, economy and other social sciences \cite{maynard_82,hofbauer_98,nowak_06}. In these systems we have an extremely large freedom in the definition of models giving the set of strategies (states or species), the interaction (payoff matrix), the connectivity structure (varied from lattice to scale-free network), and dynamical rules. Due to the large number of possibilities the complete exploration of these systems requires a long time because we need to determine separately the effect of all the mentioned ingredients of the model on the system behavior. To overcome this difficulty, the introduction of dynamical rules when not only the strategy changes but also other individual feature of players \cite{ho_jet07} may reveal the relevant region of parameter space that is important to study. During simultaneous evolutions (briefly co-evolution) of variables the success-driven Darwinian selection can serve as a general tool to identify the characteristic dynamical rules. The goal of this letter to demonstrate that the fixation of a crucial parameter is possible and the resulting value is in close connection to the state where cooperation is the largest that can be achieved at the corresponding payoff elements and topology.

The systematic investigation of the evolutionary Prisoner's Dilemma (PD) games has attracted a considerable effort in the last decades because these models can describe the ways how the cooperative behavior is maintained among selfish individuals \cite{nowak_s06}. Originally the PD is a two-person one-shot game \cite{hofbauer_98,nowak_06} where the players have two options (cooperation or defection) to choose and their income depends on their choices. The rank of possible payoffs enforces both (intelligent and selfish or rational) players to choose defection yielding the second worst income for each while the mutual cooperation provides higher income for both players. The situation is changed drastically in the multi-agent systems where the player's income comes from repeated games with the neighbors defined by a connectivity structure (lattice with nearest neighbor connections or other graphs). The evolutionary games are the combination of the multi-agent repeated games and Darwinian selection. Namely, sometimes the players are allowed to modify their strategy by imitating one of the more successful neighbors (in biological context: an offspring of the more successful species will be substituted for a less successful one). 

It turned out that the cooperative behavior can be sustained among the spatially arranged players with local interactions \cite{nowak_ijbc93} for a wide range of evolutionary (here imitation) rules even if they can follow only one of the two simplest strategies: unconditional cooperation or defection. Subsequent investigations have clarified the main effect of payoffs, connectivity structure (including networks with inhomogeneous degree distribution), and noise on the level of cooperation (for a survey see \cite{nowak_06,szabo_pr07}). These investigations highlighted a mechanism supporting the cooperation efficiently in more realistic models where the number of neighbors varies within a wide region \cite{santos_prl05,santos_prslb06} or the individuals have different personal strategy pass capability to help the imitation of their own strategy \cite{wu_pre06,szolnoki_epl07}. Evidently, the enhancement of the strategy set (e.g., the application of stochastic reactive strategies \cite{nowak_aam90}, deterministic strategies of finite memory \cite{lindgren_pd94}, and Q-learning strategies \cite{wang_plos08}) opened further dimensions towards the ways supporting cooperation.

The simultaneous evolution  (henceforth co-evolution) of strategies and another feature of the model was investigated previously by many authors. First the co-evolution of strategy distribution and connectivity structure was studied (for examples see  \cite{zimmermann_pre04,pacheco_prl06,pacheco_jtb06,biely_pd07,szolnoki_epl08,poncela_plos08,fu_pre08}). The co-evolution of the strategy distribution and inhomogeneous capability of strategy transfer was also investigated in the last years \cite{szolnoki_njp08,szolnoki_epjb09}. In another model the individuals were allowed to have different payoff matrices that can be adopted (imitated) together with the strategy, too \cite{worden_jtb07,fort_epl08}. Very recently, van Segbroeck {\it et al.} \cite{vansegbroeck_bmceb08,vansegbroeck_prl09} have introduced a co-evolutionary PD game where the players are capable to modify their connection in different ways and in parallel with the strategy adoption they can also imitate the neighbor's method used later in the rearrangement of their own neighborhood. Finally we have to mention that the co-evolution of strategy and individual learning (evolutionary) rule was investigated previously for some cases \cite{harley_jtb81,kirchkamp_jebo99,camerer_e99}. For example, Moyano and Sanchez \cite{moyano_qb08} have studied the cases when the players adopt the strategy and dynamical rule from the better player if two strategies and two rules are allowed.  

Now we extend a previous model \cite{szabo_pre05} to study what happens when the players can adopt not only the more prosperous strategy but the way of strategy adoption as well. The present set of strategy imitation rules is based on pairwise comparison of payoffs between two neighboring players chosen at random. We assume that initially the players use different rules giving the probability of strategy adoption as a function of payoff difference divided by an individual parameter resembling the temperature in the Fermi-Dirac distribution function. It will be shown that the suggested co-evolutionary process drives the system towards a final state where all the players use the same imitation rule even if their strategies are different. The state characterized by the fixed selection (learning) rule is close to the highest cooperativity (optimum) state that can be achieved applying the corresponding payoff elements and topology. As the optimum level of cooperation depends on the connectivity structure \cite{szabo_pre05,vukov_pre06} therefore our investigation is performed on both the square and kagome lattices representing two different classes of behaviors. These systems will be investigated by Monte Carlo (MC) simulations and an extended version of the dynamical mean-field approximation.

In the present model the players located on the sites $x$ of a two dimensional lattice can follow either unconditional cooperation or defection strategies, in short, $s_x=C$ or $D$. The players' income ($P_x$) come from one-shot games with the four nearest neighbors. Following the suggestion of Nowak and May \cite{nowak_n92a} we use a re-scaled payoff matrix of the so-called weak PD game, i. e., the cooperative player receives 1 or 0 if the co-player follows $C$ or $D$ strategies while the defective player is rewarded by $b$ ($1<b<2$) or 0 if the opponent cooperates or defects. Initially, each player follows a strategy ($s_x=C$ or $D$) chosen at random. Besides it we assume that the players use different imitation rules characterized by a parameter $K_x$ chosen randomly from a set of possible values $\{K_1, \ldots , K_n\}$ (as it will be detailed later on). In each subsequent elementary step of the evolutionary process we choose two neighboring players ($x$ and $y$) at random, we determine their payoff $P_x$ and $P_y$, and player $x$ adopts the strategy $s_y$ and imitation rule (characterized by $K_y$) with a probability  
\begin{equation}
W =\frac{1}{1+\exp[(P_x-P_y)/K_x]}\,
\label{eq:prob}
\end{equation}
in two (independent) consecutive processes. More precisely, we generate two random numbers ($0<r_1, r_2<1$), and $s_x \to s_y$ if $r_1 <W$ and $K_x \to K_y$ if $r_2 <W$. This means that probably both the strategy and imitation rule are adopted if $P_y - P_x \gg K_x$. Evidently, there exist elementary steps when either $s_y$ or $K_y$ or none is adopted. As a consequence of independent processes the imitation of the imitation rule is possible even if the strategies are the same ($s_x=s_y$). These dynamical rules imply the existence of absorbing states with uniform strategies and/or rules where the evolution is stopped separately. We should note, however, that qualitatively similar results were observed when imitation of rules was only possible if players have different strategies. 

The individual parameter $K_x$ of player $x$ can be interpreted in different ways \cite{blume_geb03,szabo_pr07}. On the one hand we can think that in realistic systems the payoff matrix describes the average payoff and the current payoffs should be modified by a stochastic term as it is modelled by Perc \cite{perc_njp06c} and Traulsen {\it et al.} \cite{traulsen_jtb07}. The noisy term can be caused by the fluctuating environment, cognitive mistakes, etc. For a suitable probability distribution of the stochastic contribution, the deterministic imitation of the better player can yield a strategy adoption rule similar to those given by (\ref{eq:prob}). In that case $K_x$ characterizes the amplitude of noise. On the other hand, the personal decision of players can also involve stochastic elements reflecting their freedom to not accept the better strategy or even to follow the worse one (for the latter interpretation $K_x$ denotes the average amount of payoff what player $x$ hazards when looking for a better solution).

The evolutionary process is governed by repeating the mentioned elementary steps that drive the system towards a final state described by the average portion $\rho$ of cooperators and the distribution of $K_x$. If initially the players use a uniform rule ($K_x=K$, for $\forall x$) then this system becomes equivalent to those studied previously \cite{szabo_pre05}. In that case we can distinguish three regions of $b$ dependent on $K$. If $b<b_{c1}(K)$ then only cooperators remain alive after a transient period. On the contrary, only defectors will survive in the final state when $b>b_{c2}(K)$. Within the intermediate region [$b_{c1}(K)< b < b_{c2}(K)$]  the stationary value of $\rho$ decreases from 1 to 0 if $b$ is increased.

First we study the present model on the square lattice where both $b_{c1}(K)$  and $b_{c2}(K)$ goes to 1 if $K$ tends to either zero or infinity. Besides it, there exists an optimum value of $K$ where $b_{c2}(K)$ reaches its maximum. A one-peak profile (for an example see Fig. 31 in \cite{szabo_pr07}) can be observed when evaluating $\rho$ as a function of the homogeneous $K$ for fixed $b$ if $1 < b < \max [b_{c2}(K)]$. In the latter case we can introduce two threshold values of $K$ in a way that cooperators die out if $K<K_{c1}(b)$ or $K>K_{c2}(b)$. Within the intermediate region of $K$ [$K_{c1}(b) < K < K_{c2}(b)$] the $C$ and $D$ strategies coexist for the given uniform rule.

As the above investigations \cite{szabo_pre05} have also indicated that the relaxation time diverges if $K \to 0$ or $\infty$ therefore the undesired consequences of this effect was avoided by introducing additional constraints, namely, all $K_i>K_{\rm min}$ (typically $K_{\rm min}=0.001$). On the other hand, several runs have justified that rules with high $K_x$ die out fast, therefore the initial set of $K_i$ has also been limited from above (typically $K_{\rm max}=2$) and $n$ is varied from 2 to 200 for sake of simplicity.

Let us discuss the trivial situations when the players have different $K_x$ parameters in the initial state but their value exceeds the second threshold value, that is $K_x > K_{c2}(b)$ for $\forall x$. After some time only defectors remain alive ($s_x=D$) with a preference of lower $K_i$. When cooperator strategies become extinct all players receives the same payoff, $P_x=0$, and the further evolution of rules ($K_x$) can be well described by the voter model (for a survey see \cite{liggett_85}) with a large number of candidates. This means that one can observe growing domains of players with the same rules and the typical domain size increases with the logarithm of time in the two-dimensional systems. The same phenomenon is found if $K_x < K_{c1}(b)$ for $\forall x$ as well as for the combination of the latter two cases when there is no $K_x$ within the intermediate region [$K_{c1}(b) < K_x < K_{c2}(b)$] in the initial state.

\begin{figure}
\scalebox{0.5}[0.5]{\includegraphics{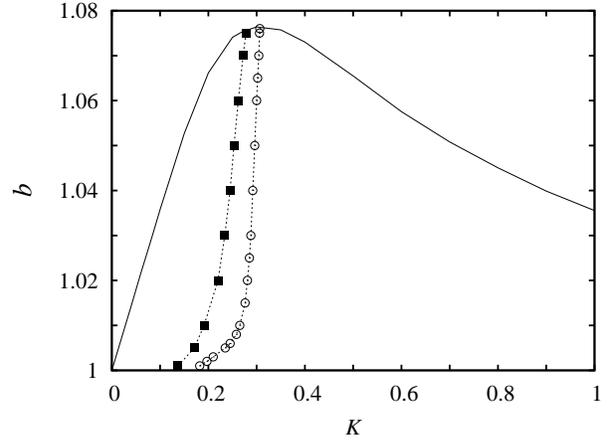}}
\caption{The MC results on the square lattice for the fixation values are denoted by closed squares within the coexistence region bounded by the solid line ($b_{c2}(K)$). Open circles show the position of local maximum in the $\rho$ portion of cooperators. Dotted lines are just to guide the eye.}
\label{fig1}
\end{figure}

The final state of the co-evolutionary process changes drastically if initially there are several players with imitation rules belonging to the 
coexistence region for the homogeneous cases, i.e., $K_{c1}(b) < K_x < K_{c2}(b)$. The MC simulations have indicated clearly that after a relaxation period all the players use the same imitation rule. The Darwinian selection chooses the rule $K_i \in \{K_1, \dots, K_n\}$ that has the ''minimum distance'' from a fixation value $K_f(b)$. The quotation mark refers to a possible asymmetry between the two sides, however, the estimation of its magnitude is prevented by the statistical error. Apparently the Darwinian selection favors a rule $K_i$ providing the highest average payoff (as it occurs for population dynamics) here, however, the value of $K_f(b)$ does not coincide the values of $K$ exhibiting local maximum in $\rho$ or average payoff (in general the difference between the latter two quantities is smaller than our statistical error comparable to symbol size). Figure \ref{fig1} demonstrates the fixation values within the coexistence region and also the position of local maximum of $\rho$ used frequently to quantify the cooperativity in the whole society.

Naturally, the fixation time depends on the system size $L$, the initial set of $K_i$ values, and also the number ($n$) of different values.  It turned out that for sufficiently large system sizes ($200 < L$) the selected rule becomes independent of the initial configuration and sequence of random numbers. The efficiency of the accurate determination of $K_f(b)$ could be improved significantly if only two rules were allowed in the initial state as detailed below. 

As mentioned above the topological feature of the connectivity structure influences the qualitative behavior (phase diagram) in the evolutionary PD games \cite{szabo_pre05}. On the kagome lattice overlapping triangles support the spreading of cooperative behavior in the low noise limit. For homogeneous imitation rules the upper boundary of the coexistence region ($b_{c2}(K)$) decreases monotonously from $3/2$ to 1 if $K$ increases from 0 to $\infty$ \cite{vukov_pre06}. This behavior implies the possibility that here the Darwinian selection of rules (within the coexistence region) favors the lowest values of $K_i$ referring to $K_f(b)=0$. This behavior has indeed been justified by MC simulations if $b$ exceeds a threshold value ($b>b_{\rm th}=1.182(2)$). For low values of $b$ we have found a behavior resembling those observed on square lattice.

\begin{figure}
\scalebox{0.5}[0.5]{\includegraphics{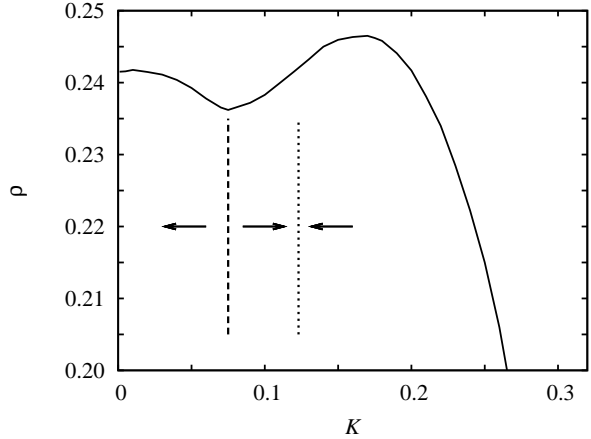}}
\caption{Portion of cooperators as a function homogeneous $K_x=K$ for $b=1.17$ on the kagome lattice. The MC results are illustrated by a solid line because the statistical error is comparable to the line thickness. The horizontal arrows indicate that the system can evolve towards the fixation value $K_f(b)$ (denoted by dotted vertical line) through weak mutations. At the same time the weak mutations drives the system towards a state where $K_x \to 0$ if initially each $K_x$ is smaller than a threshold value denoted by the dashed vertical line.}
\label{fig2}
\end{figure}

Figure \ref{fig2} shows the $K$-dependence of $\rho$ (for homogeneous rules $K_x=K$ if $b=1.17$) in a magnified plot to emphasize the existence of two local maxima separated by a shallow local minimum. If the co-evolutionary system is started from a state with many rules inside the coexistence region then only one rule (the corresponding $K_f(b)$ is denoted by the vertical dotted line in Fig. \ref{fig2}) will remain alive in a way as described above. There exists, however, a relevant difference in the behaviors between the square and kagome lattices. Namely, on the kagome lattice two attractors (final imitation rules) can be observed. The horizontal arrows in Fig.~\ref{fig2} illustrate the direction of preference if initially the players follow rules from the marked intervals. The result of these types of investigations can be interpreted as the direction of evolution in $K_x$ through rare and weak mutations. Although the state of $K_x=0$ $\forall x$ has a finite basin of attraction through a weak mutation this state is overcome by the offspring of players of $K_f(b)$ being present initially.  

\begin{figure}
\scalebox{0.5}[0.5]{\includegraphics{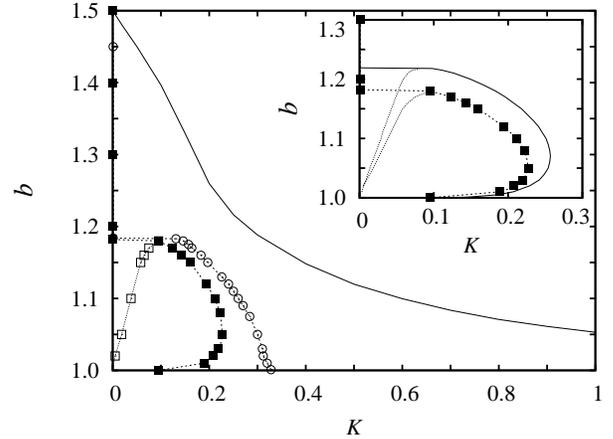}}
\caption{Fixation values $K_f(b)$ for the competing imitation rules on the kagome lattice are illustrated by closed squares. Open squares indicate the position of separatrix indicated by dashed line in Fig.~\ref{fig2}. Solid line represents the maximum values of $b$ where cooperators can survive for homogeneous rules. Open circles show the position of local maxima of $\rho$ within the coexistence region. The insert compares the prediction (solid line) of an extended version of the three-site dynamical cluster method with the results (symbols) of MC simulations.}
\label{fig3}
\end{figure}

The MC results for arbitrary values of $b$ are summarized in Fig.~\ref{fig3} where the cases of $K_f(b)\simeq 0$ are denoted by several closed squares positioned at $K_{\rm min}$ (instead of 0) used to avoid the above mentioned technical difficulties. In these cases the MC simulations have indicated a plateau (within the statistical error) in the values of $\rho$ and average payoff. If the value of $b$ is decreased gradually then an abrupt change of $K_f(b)$ is found at $b=b_{\rm th}$. Below this threshold value there appears a positive $K_f(b)$ that can also be related to the local maxima in the portion of cooperators. Notice that $K_f(b)$ correlates weakly with the position of the second (right) local maximum of $\rho$ (see Fig. \ref{fig2}) if $b<b_{\rm th}$. The height of the second local maximum decreases monotonously if $b$ is increased and this local peak vanishes above a value larger than $b_{\rm th}$. 

As mentioned, the selected rule ($K_f(b)$) can be determined more efficiently if we consider the competitions between only two suitable imitation rules. This approach can also be utilized in the extended version of dynamical cluster techniques (for a brief survey see \cite{szabo_pr07}) where we derive a set of equations of motion for the probability of each (strategy and rule) configuration existing on a given cluster of sites. The accuracy of this method can be improved by choosing larger clusters. Previous investigations \cite{szabo_pre05} have justified that the three-site (triangular) cluster of the kagome lattice is the smallest one that gives adequate description about all the relevant features for homogeneous rules. This fact has raised the possibility to extend this technique for the two-rule cases. The details of this method will be published elsewhere, now we only compare its prediction with the MC results in the insert of Fig.~\ref{fig3}. Noteworthy that this method predicts a little bit higher threshold value for the payoff parameter $b$, namely, $b_{\rm th}^{\rm (3s)}=1.219(2)$, and confirms the difference between the selected rule and local maxima both in $\rho$ and average payoff.

In summary, the Darwinian selection (imitation of the better) proved to be beneficial for the whole society for the Prisoner's Dilemma if not only the strategy but also the way of strategy adoption is adopted from a successful neighboring player. The systematic investigations highlight the relevance of the selected dynamical rules that, depending on the connectivity structure and payoff, provides the highest or almost the highest possible average income. The small difference between the selected and the optimal dynamical rules might have been related to the spatial effects enhancing the importance of fluctuations.

\acknowledgments
This work was supported by the Hungarian National Research Fund
(Grant No. K-73449).


\begin{thebibliography}{0}

\bibitem{maynard_82}
  \Name{Maynard~Smith J.}
  \Book{Evolution and the theory of games}
  \Publ{Cambridge University Press, Cambridge}
  \Year{1984}.

\bibitem{hofbauer_98}
  \Name{Hofbauer J. \and Sigmund K.}
  \Book{Evolutionary Games and Population Dynamics}
  \Publ{Cambridge University Press, Cambridge}
  \Year{1998}.

\bibitem{nowak_06}
  \Name{Nowak M.~A.}
  \Book{Evolutionary Dynamics: Exploring the Equations of Life}
  \Publ{Harvard University Press, Cambridge, MA}
  \Year{2006}.

\bibitem{ho_jet07}
  \Name{Ho T.~H., Camerer C.~F. \and Chong J.-K.}
  \REVIEW{J. Econ. Theory.}{133}{2007}{177}.

\bibitem{nowak_s06}
  \Name{Nowak M.~A.}
  \REVIEW{Science}{314}{2006}{1560}.

\bibitem{nowak_ijbc93}
  \Name{Nowak M.~A. \and May R.~M.}
  \REVIEW{Int. J. Bifurcat. Chaos}{3}{1993}{35}.

\bibitem{szabo_pr07}
  \Name{Szab{\'o} G. \and F{\'a}th G.}
  \REVIEW{Phys. Rep.}{446}{2007}{97}.

\bibitem{santos_prl05}
  \Name{Santos F.~C. \and Pacheco J.~M.}
  \REVIEW{Phys. Rev. Lett.}{95}{2005}{098104}.

\bibitem{santos_prslb06}
  \Name{Santos F.~C., Rodrigues J.~F. \and Pacheco J.~M}
  \REVIEW{Proc. Roy. Soc. Lond. B}{273}{2006}{51}.

\bibitem{wu_pre06}
  \Name{Wu Z.-X., Xu X.-J., Huang Z.-G., Wang S.-J. \and Wang Y.-H.}
  \REVIEW{Phys. Rev. E}{74}{2006}{021107}.

\bibitem{szolnoki_epl07}
  \Name{Szolnoki A. \and Szab{\'o} G.}
  \REVIEW{Europhys. Lett.}{77}{2007}{30004}.

\bibitem{nowak_aam90}
  \Name{Nowak M.~A. \and Sigmund K.}
  \REVIEW{Acta Appl. Math.}{20}{1990}{247}.

\bibitem{lindgren_pd94}
  \Name{Lindgren K. \and Nordahl M.~G.}
  \REVIEW{Physica D}{75}{1994}{292}.

\bibitem{wang_plos08}
  \Name{Wang S., Szalay M.~S., Zhang Z. \and Csermely P.}
  \REVIEW{PLoS ONE}{3}{2008}{e1917}.

\bibitem{zimmermann_pre04}
  \Name{Zimmermann M.~G., Egu{\'{\i}}luz V. \and San~Miguel M.}
  \REVIEW{Phys. Rev. E}{69}{2004}{065102(R)}.

\bibitem{pacheco_prl06}
  \Name{Pacheco J.~M., Traulsen A. \and Nowak M.~A.}
  \REVIEW{Phys. Rev. Lett.}{97}{2006}{258103}.

\bibitem{pacheco_jtb06}
  \Name{Pacheco J.~M., Traulsen A. \and Nowak M.~A.}
  \REVIEW{J. Theor. Biol.}{243}{2006}{437}.

\bibitem{biely_pd07}
  \Name{Biely C., Dragosits K. \and Thurner S.}
  \REVIEW{Physica D}{228}{2007}{40}.

\bibitem{szolnoki_epl08}
  \Name{Szolnoki A., Perc M. \and Danku Z.}
  \REVIEW{Europhys. Lett.}{87}{2008}{50007}.

\bibitem{poncela_plos08}
  \Name{Poncela J., G{\'o}mez-Garde{\~n}es J., Flor{\'{\i}}a L.~M., Sanchez A. \and Moreno Y.}
  \REVIEW{PLoS ONE}{3}{2008}{e2449}.

\bibitem{fu_pre08}
  \Name{Fu F., Hauert C., Nowak M.~A. \and Wang L.}
  \REVIEW{Phys. Rev. E}{78}{2008}{026117}.

\bibitem{szolnoki_njp08}
  \Name{Szolnoki A. \and Perc M.}
  \REVIEW{New J. Phys.}{10}{2008}{043036}.

\bibitem{szolnoki_epjb09}
  \Name{Szolnoki A. \and Perc M.}
  \REVIEW{Eur. Phys. J. B}{67}{2009}{337}.

\bibitem{worden_jtb07}
  \Name{Worden L. \and Levin S.~A.}
  \REVIEW{J. Theor. Biol.}{245}{2007}{411}.

\bibitem{fort_epl08}
  \Name{Fort H.}
  \REVIEW{Europhys. Lett.}{81}{2008}{48008}.

\bibitem{vansegbroeck_bmceb08}
  \Name{Van~Segbroeck S., Santos F.~C., Now{\'e} A., Pacheco J.~M. \and Lenaerts T.}
  \REVIEW{BMC Evol. Biol.}{8}{2008}{287}.

\bibitem{vansegbroeck_prl09}
  \Name{Van~Segbroeck S., Santos F.~C., Lenaerts T. \and  Pacheco J.~M.}
  \REVIEW{Phys. rev. Lett.}{102}{2009}{058105}.

\bibitem{harley_jtb81}
  \Name{Harley C.~B.}
  \REVIEW{J. Theor. Biol.}{89}{1981}{611}.

\bibitem{kirchkamp_jebo99}
  \Name{Kirchkamp O.}
  \REVIEW{J. Econ. Behav. Org.}{40}{1999}{295}.

\bibitem{camerer_e99}
  \Name{Camerer C. \and Ho T.-H.}
  \REVIEW{Econometrica}{67}{1999}{827}.

\bibitem{moyano_qb08}
  \Name{Moyano L.~G. \and S{\'a}nchez A.}
  \REVIEW{J. Theor. Biol.}{}{2009}{doi:10.1016/j.jtbi.2009.03.002}.

\bibitem{szabo_pre05}
  \Name{Szab{\'o} G., Vukov J. \and Szolnoki A.}
  \REVIEW{Phys. Rev. E}{72}{2005}{047107}.

\bibitem{vukov_pre06}
  \Name{Vukov J., Szab{\'o} G. \and Szolnoki A.}
  \REVIEW{Phys. Rev. E}{73}{2006}{067103}.

\bibitem{nowak_n92a}
  \Name{Nowak M.~A. \and Sigmund K.}
  \REVIEW{Nature}{355}{1992}{250}.

\bibitem{blume_geb03}
  \Name{Blume L.~E.}
  \REVIEW{Games Econ. Behav.}{44}{2003}{251}.

\bibitem{perc_njp06c}
  \Name{Perc M.}
  \REVIEW{New J. Phys.}{8}{2006}{183}.

\bibitem{traulsen_jtb07}
  \Name{Traulsen A., Nowak M.~A., and Pacheco J.~M.}
  \REVIEW{J. Theor. Biol.}{244}{2007}{349}.

\bibitem{liggett_85}
  \Name{Liggett T.~M.}
  \Book{Interacting Particle Systems}
  \Publ{Springer-Verlag, New York}
  \Year{1985}.

\end{thebibliography}
\end{document}